\documentclass[aps,twocolumn,showpacs,amsmath,amssymb,floatfix]{revtex4}

\usepackage{graphicx}
\usepackage{amssymb}
\usepackage{subfigure}
\usepackage{dcolumn}
\usepackage{bm}

\begin{document}
\title{Characterisation of spatial network-like patterns from junctions' geometry.} 
\author{Andrea Perna$^{1,2,4}$,  Pascale Kuntz$^{2}$, St\'ephane Douady$^{3}$}

\affiliation{
$^1$ Institut des Syst\`{e}mes Complexes Paris \^Ile-de-France, 57-59 rue Lhomond, F-75005, Paris, France \\
$^2$ \'{E}cole Polytechnique de l'Universit\'{e} de Nantes, 19 rue Christian Pauc, BP50609, 44306 Nantes, France \\
$^3$ Laboratoire Mati\`{e}re et Syst\`{e}mes Complexes UMR 7057,
B\^atiment Condorcet Universit\'e Paris Diderot - CC7056, 75205 Parix cedex 13, France \\
$^4$ Mathematics Department, Uppsala University, 75106 Uppsala, Sweden}

\begin{abstract}
We propose a new method for quantitative characterization of spatial network-like patterns with loops, such as surface fracture patterns, leaf vein networks and patterns of urban streets. Such patterns are not well characterized by purely topological estimators: also patterns that both look different and result from different morphogenetic processes can have similar topology. 
A local geometric cue -the angles formed by the different branches at junctions- can complement topological information and allow to quantify the large scale spatial coherence of the pattern. For patterns that grow over time, such as fracture lines on the surface of ceramics, the rank assigned by our method to each individual segment of the pattern approximates the order of appearance of that segment.
We apply the method to various network-like patterns and we find a continuous but sharp dichotomy between two classes of spatial networks: \textit{hierarchical} and \textit{homogeneous}. The first class results from a sequential growth process and presents large scale organization, the latter presents local, but not global organization.

\end{abstract}

\maketitle

\section{Introduction}
A central issue in complex systems research is to understand the formation and the properties of spatio-temporal patterns found in physics and biology. To this end, an essential step is the definition of appropriate measures of their form: quantitative estimators that allow to compare different patterns and to objectively validate models. This paper focuses on the family of spatial patterns that are ``network-like''.

Network-like patterns are common in natural and artificial systems: they are found in leaf veins, fractures on the surface of materials, patterns of urban streets and animal trails~/~galleries, river networks, blood vessels and circulatory systems. The factors underlying the formation of these patterns are different from system to system: surface cracks result from the shrinkage and stress of materials; urban streets are generated by human activity and so on. In spite of intrinsic differences, a few simple morphogenetic events describe the formation of all these patterns: nucleation of new network components, elongation of existing segments, branching and intersection. The final topology of the pattern is completely determined by the sequence of such growth events, plus a pruning event: the cut or removal of already formed segments.

Many studies have used topological estimators to describe the form of network-like patterns and better understand their morphogenesis and functional properties. One of the first progresses in this direction goes back to the Horton-Strahler coefficients~\cite{Horton1945,Strahler1952}, a simple but powerful tool to describe quantitatively the form of hydrogeologic networks. The coefficients, which work by assigning rank numbers to every branch of a tree-like network, were successfully applied to the study of different systems, from river networks~\cite{Dodds2000, Dodds2000a, Dodds2000b}, to leaf patterns~\cite{Pelletier2000}, and even to ant trail~\cite{Ganeshaiah1991} and termite gallery~\cite{Hedlund1999} patterns. Unfortunately, Horton--Strahler's coefficients cannot be computed on networks with cycles. This has a profound impact on the applicability of the method, as many real-world network-like patterns have cycles.

General graphs are characterized by a whole set of measures of the local and global organization: node degree, assortativity between nodes, clustering coefficient, frequency of specific subgraphs, presence and number of cycles, diameter, path length etc.~\cite{Boccaletti2006,Costa2007,Barthelemy2010}. Several studies have computed graph measures and used them to quantify the local and large scale organization of real world spatial patterns, such as urban street patterns~\cite{Buhl2006a,Crucitti2006a,Scellato2006,Cardillo2006,Porta2006,Barthelemy2008,Jiang2004,Travencolo2008,Gastner2006b}, systems of animal trails~\cite{Buhl2009} and galleries~\cite{Buhl2004,Perna2008,Perna2008a,Perna2008b,Valverde2009}, networks of channels in trabecular bones~\cite{Costa2006,Viana2009}, networks of fungal hyphae~\cite{Bebber2007}.

Unfortunately, purely topological estimators, however useful, do not fully account for the organization of 2D network-like patterns which undergo specific constraints. For instance, in planar graphs, the average node degree cannot be higher than six~\cite{Barthelemy2010}. In addition, for a number of network-like patterns, the degree distribution is even more regular than what is imposed by planarity: the large majority of nodes is found to have degree equal to three in very different systems, such as two-dimensional foams~\cite{Weaire1999}, honeycomb patterns and biological epithelia~\cite{Thompson1992}, leaf vein networks~\cite{Bohn2002}, fracture patterns~\cite{Bohn2005}. In self-organized, ``bottom-up'' towns~\cite{Buhl2006a,Cardillo2006} the majority of junctions between streets also have degree three (though the same does not hold for planned towns, where the majority of crossroads have degree four~\cite{Cardillo2006} or even higher~\cite{Lammer2006}).  The homogeneity of degree is reflected by an homogeneity in the length of network cycles that are composed by six edges on average in all the aforementioned systems. Similarly, network distances are not very informative as they basically scale with euclidian distances in all these systems.

Here we want to use the spatial information to provide a deeper understanding of the pattern. We focus in particular on the information carried by the angles formed at junctions. This choice is motivated by two complementary arguments, one about the physics of the growth of the network-like pattern, which can influence directly the branching angles, and the other about dynamic properties (e.g. navigation) that may be associated with the resulting network.

Throughout the text, as we consider spatial networks, we will make extensive use of the words \textit{edge}, \textit{arc} and \textit{segment} with the following definitions: an \textit{edge} is a curve -generally almost linear- between two consecutive junctions (classical network edge); an \textit{arc} is a directed edge when an orientation is added to the network; a \textit{segment} is a series of contiguous edges grouped together as described later in section \ref{sec:algorithm} and used to reveal a structuration at a larger scale.

\subsection{growth}
\label{sec:growth}

\begin{figure}[!tbp]
\vspace{0cm} 
\resizebox{1\columnwidth}{!}{
	\includegraphics{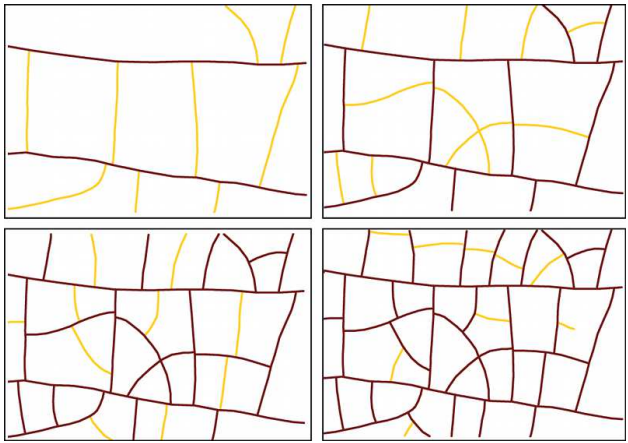}
}
	\caption{(Color online) Drawing of the evolution of a fracture pattern in a solution of latex particles~\cite{Bohn2005a}. In each image, the yellow (light gray) lines indicate the newly appeared fractures. When new lines of fracture appear their growth is affected by the already existing fractures (for instance, newer fractures do not cross older ones). Conversely, the position of older lines does not change when new ones appear.}
	\label{fig:examplecrack}
\end{figure}

Network-like patterns often are formed as the result of a sequential process, where new segments appear at different times~\cite{Bohn2005} without undergoing further reorganization. One example of such patterns is provided by the fracture lines on the glaze of ceramics illustrated in figure~\ref{fig:examplecrack}.
Bohn et al.~\cite{Bohn2005} suggested that urban streets patterns fall into this same morphological class. 

The appearance, elongation and termination of new segments is affected by the older segments; the older segments, however, do not undergo further reorganisation as a consequence of the appearance of new ones. Under these conditions (sequential process, growth with no deletion of edges and absence of reorganization) the temporal hierarchy in the appearance of segments is reflected into the spatial hierarchy of lengths and arrangements of the final pattern~\cite{Bohn2005a}. Across a junction the edges belonging to the older segment are the straight continuation of one another; conversely, the edge belonging to the newer segment typically forms a large angle with the others. 

\begin{figure}[b!]
\vspace{0cm} 
\resizebox{1\columnwidth}{!}{
	\includegraphics{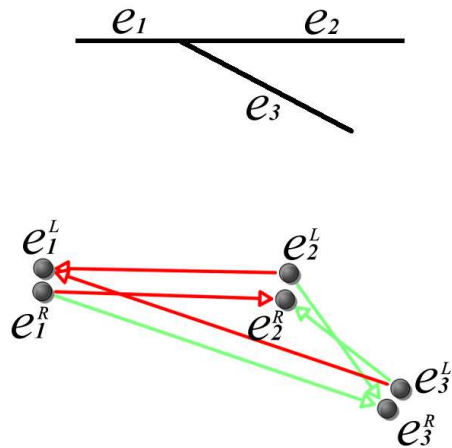}
}
	\caption{(Color online) \textbf{A}: Small graph, involving a single junction. \textbf{B}: The weighted directed line graph built from the graph in A. Weights are as follows: red (dark gray)~$\rightarrow$~weight=0; green (light gray)~$\rightarrow$~weight=1}
	\label{fig:algorithm}
\end{figure}
But the local angle information is not sufficient in itself to attribute the edges to the same or a different segment: consider the case of a junction like the one in figure \ref{fig:algorithm}-A. In the absence of any information, the most natural inference is to group edges $e_1$ and $e_2$ into the same, older, segment. The attribution is more problematic if one already knows that $e_3$ is older: then it would be more natural to group $e_3$ and $e_1$ into the same old segment, and assume $e_2$ appeared later.
In summary, the best inference about what network edges should be grouped into the same segment depends on the continuity between adjacent edges, but also on the information already available about the order of appearance of other segments. It can be constructed only sequentially.

\subsection{navigation}
\label{sec:navigation}

When the network-like pattern is also the support for a transportation function, as it is for instance the case with patterns of urban streets, the angles of edges at intersections and junctions play a role for navigation and orientation. 

In architecture, this concept has been theorized under the name of ``space syntax''~\cite{Hillier1984}. With the space syntax method of analysis, urban street patterns are fragmented into a number of straight segments (ideally the minimum number of segments in which the line of view is conserved) and one analyses the network obtained representing straight segments as nodes, and intersections between segments as edges. Then selecting a line (a street) as a starting point, one can number each line in the map according to how many changes of direction separate it from the starting line. This measure is generally referred to as ``depth'' and is a measure of distance: it represents the minimum number of changes of direction to go from the origin to any other place in the street network. This kind of measure has been correlated with different aspects of social life: patterns of pedestrian movement, traffic, property value and so on~\cite{Hillier1993}. Nevertheless, the process of fragmenting street patterns into segments with the same line of sight was not proven to always have a computable solution and it might be too sensitive to small differences of orientation of the urban grid~\cite{Ratti2004}.

In the physics literature, analogous ideas are followed through what is usually called a ``dual network'' approach (though a more appropriate definition would be ``line network'' approach), where named streets~\cite{Jiang2004,Rosvall2005,Kalapala2006} or contiguous segments~\cite{Porta2006} are mapped into network nodes and there is an edge whenever two streets intersect or bifurcate. Dual networks also provide a convenient representation for measuring the amount of information necessary to navigate inside the network~\cite{Rosvall2005}, in particular for navigation strategies relying more upon the continuity of linear elements than on salient points.

In fact one can imagine to describe a path through the network with instructions of the form: "go straight for N steps (N junctions or crossroads), or until you find a salient point (a traffic light, a square...)", followed by information on which new direction to take. In general a ``simple'' path will be one that involves only few deviations from the current direction, not necessarily a short one. However, the result will not be the same depending on the direction followed.  Let's illustrate this with reference to the junction of figure \ref{fig:algorithm}-A. Moving from edge $e_1$ to $e_2$, or in the opposite direction from $e_2$ to $e_1$ does not involve any change of direction. Going from $e_1$ to $e_3$ requires one change of direction (i.e. to abandon the straightest path and take an alternative one), but the opposite is not true since $e_1$ is the more direct continuation of movement when coming from $e_3$.

Both the argument about network growth and the argument about navigation illustrate the interest of classifying the edges of network-like patterns using information provided by junctions' angles. Our exaxmples also indicate that the process is not symmetric with respect to the starting point used for classification. 

In the following section \ref{sec:algorithm} we introduce a new algorithm that given a local rule (namely an evaluation of the angles at each junction) and a set of arbitrarily chosen root edges, assigns a rank (expressed by an integer number) to each edge of the network-like pattern. Contiguous edges with the same rank can then be grouped together into segments. The numbers of segments for each given rank and their average properties provide a quantitative statistical characterization of the morphology of the pattern. Possible applications to lattices and real world spatial networks are explored in the remaining sections \ref{sec:lattices} to \ref{sec:conclusions}.

\section{Algorithm description}
\label{sec:algorithm}

The classification and numbering of segments involves two steps. First, given a root edge, or a set of root edges, a rank number is assigned to each edge. Then, the contiguous edges with the same rank are grouped into segments.

\subsection{computing edge ranks}
\label{subsec:edgeranks}

Let us denote by $G$ a graph which models a network-like structure. Given a root edge $e_{r}$ in $G$, the rank of an edge $e_{i}$ of $G$ expresses the minimum number of direction changes needed to reach $e_{i}$ when coming from $e_{r}$. We can also take into account a set $S_{r}$ of root edges; in this case, the rank of $e_{i}$ is the minimum number of direction changes when coming from the closest root edge in $S_{r}$.

The propagation of ranks across a junction depends on the direction in which the junction is traversed. For instance, in the graph of figure \ref{fig:algorithm}-A if we set the rank of edge $e_{1}$ to zero ($r(e_{1}) = 0$ by convention), then the rank of  $e_{3}$ is equal to 1 ($r(e_{3}) = 1$), but the same relation does not hold in the opposite direction: if we set the rank of $e_{3}$ to zero ($r(e_{3}) = 0$ by convention), then the rank of $e_{1}$ is also equal to zero ($r(e_{1}) = 0$: no deviation from the straightest path when going from $e_{3}$ to $e_{1}$). 

Generally speaking, each edge $e_{i}$ can be crossed in two directions; by convention, we denote $e_{i}^{R}$ (resp. $e_{i}^{L}$) the arc associated to $e_{i}$ when it is crossed towards the right (resp. towards the left). To make the computation of all the ranks easier, we consider the directed line graph $G_{D}$ (fig \ref{fig:algorithm}-B) which models the connections between all the arcs. The vertices of $G_{D}$ are the arcs $\left\{ e_{i}^{R},e_{i}^{L}\right\} $ associated to any edge of $G$, and two vertices $e_{i}^{*}$ and $e_{j}^{*}$ (the star means either $R$ or $L$) are linked together in $G_{D}$ if the destination of $e_{i}^{*}$ is also the origin of $e_{j}^{*}$.

For instance, the crossing of the path $e_{1}-e_{2}$ from the left is modeled by $e_{1}^{R}\rightarrow e_{2}^{R}$ in $G_{D}$, and the crossing in the opposite direction by $e_{2}^{L}\rightarrow e_{1}^{L}$. There is no connection between $e_{1}^{L}$ and $e_{2}^{R}$ as the navigation $e_{1}^{L} \rightarrow e_{2}^{R}$ is not directly possible.

Weights are assigned to the arcs of $G_{D}$ as follows. The weight $w\left(e_{i}^{*},e_{j}^{*}\right)$ is equal to $0$ when the path formed in the initial graph by the edges $e_{i}$ and $e_{j}$ is the straightest one and $1$ otherwise. 
We can also introduce a threshold on the maximum angle, implying that if the straightest pairing of edges still involves a too large angle, then it also will have weight equal to 1. Formally, $w\left(e_{i}^{*},e_{j}^{*}\right)=0$ when (1) the angle formed in the initial graph $G$ by $e_{i}^{*}$ and $e_{j}^{*}$ is the minimum angle formed by $e_{i}^{*}$ and its adjacent edges, and (2) this angle is smaller than a given threshold ($45$ degrees in our analyses).

Once defined a root edge $e_{r}$ in the initial graph whose rank is equal to $0$ by convention, the rank $r\left(e_{i}\right)$ of any other edge $e_{i}$ can be easily obtained by a distance computation in $G_{D}$:
\begin{equation}
\begin{split} 
r\left(e_{i}\right)= \\
Min\left\{ d\left(e_{r}^{L},e_{i}^{L}\right),d\left(e_{r}^{L},e_{i}^{R}\right),d\left(e_{r}^{R},e_{i}^{L}\right),d\left(e_{r}^{R},e_{i}^{R}\right)\right\}
\end{split}
\end{equation}
where $d$ is the shortest path length in the directed line graph $G_{D}$ weighted by $w$. When a set $S_{r}$ of root vertices is considered, the rank of $e_{i}$ is the minimum of the ranks computed from all the edges of $S_{r}$ with the above formula.

\subsection{grouping edges into segments}
\label{subsec:segments}

Several adjacent edges with the same rank can be grouped together into ``segments''. A segment is a series of edges all having the same rank and with exactly two ends (not including junctions).
The segment is initialized with any edge of $G$. Its continuation is determined by pairing together adjacent edges of the same rank until one of the following conditions is encountered:
\begin{enumerate}
\item There are no more edges with the same rank at one extremity. \label{list:no_more_edges}
\item The segment intersects an edge of lower rank. \label{list:lower_rank}
\end{enumerate}

\begin{figure}[bthp]
\vspace{0cm} 
\resizebox{0.8\columnwidth}{!}{
	\includegraphics{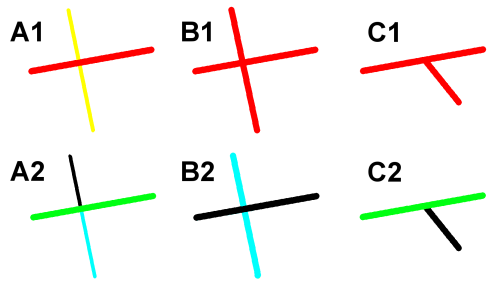}
}
	\caption{(Color online) Illustration of the procedure for grouping together edges into segments. \textbf{A1}: the red (near-horizontal) edges have lower rank than the yellow (near-vertical) edges. The edges are grouped in \textbf{A2} into three segments: a green (near-horizontal) lower rank segment and two (black and cyan) segments with higher rank. \textbf{B1}: all the edges have the same rank. In this case, two straight segments are identified in \textbf{B2}. If an odd number of segments with the same rank meet at a junction (\textbf{C1}), the one that deviates more from the direction of the others forms a distinct segment (\textbf{C2})}
	\label{fig:group_segments}
\end{figure}
In the special case when three or more edges with the same rank are incident to the same junction, the continuation of the segments is determined by pairing together the two edges that provide the straightest continuation of one another. Then, we proceed in a similar way pairing together all the other edges incident to the node. If an odd number of edges with the same rank meet at a junction, one edge remains excluded from all the pairings and the segment containing that edge is ended at the junction. 

We can give an intuitive justification for the above rules. Imagine the case of a network-like pattern resulting from a sequential growth process, such as for instance fracture lines on ceramics. Grouping edges into segments is equivalent to identify those network edges that belong to the same line of fracture. With condition \ref{list:lower_rank} we impose that younger lines of fracture (higher ranks) never cross already formed fracture lines. A parallel can also be found with what happens for urban street patterns, where small streets usually change their name when crossing larger ones.

Figure \ref{fig:group_segments} shows examples of how segments are grouped together. In A1 the near-horizontal edges have lower rank than the near-vertical edges. In this case, the near-vertical edges form two distinct segments (A2), in spite of being in direct continuity. B1 presents exactly the same pattern, except that now all the edges have the same rank. In this case, the edges are grouped together in two intersecting segments, one near-horizontal and one near-vertical (B2). When there is an odd number of edges incident to a junction, as in C1, the couple(s) of edges that deviate less from each other's direction are paired together in the same segment(s) and the remaining edge forms a segment by itself, which is ended at the junction (C2).

To summarize, the straight contiguity of edges is well represented by a directed line graph $G_D$ whose arcs are weighted with appropriately chosen weights. The arbitrary selection of a set of root edges in $G$ allows to express changes of direction in terms of a distance measure, allowing to assign each edge of $G$ a ranking number. Edges that are more likely to belong to the same spatio-temporal event of network formation can be grouped together into segments. The C++ computer code implementing the rank computation and segment mapping is available as online supplementary material, alongside with sample networks to test it.

\section{Lattice models}
\label{sec:lattices}

\begin{figure}[!t]
\vspace{0cm} 
\resizebox{0.7\columnwidth}{!}{
	\includegraphics{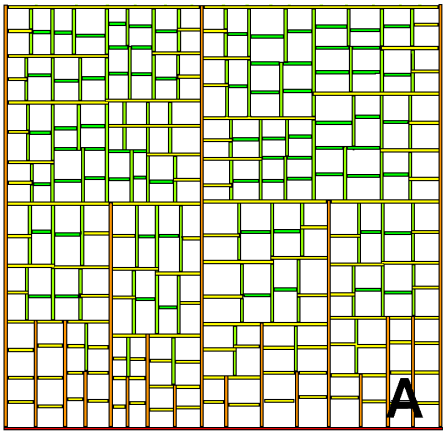}
}
	\caption{(Color online) Mondrian pattern generated by iterated domain divisions. Here, divisions are alternating: all the cuts made at division $t+1$ are orthogonal to the cuts at division $t$. In all figures, the color (shade of gray) of the edges reflects their rank; a periodic rainbow colormap is used. For this example, all the edges in the bottom of the figure are selected as roots.}
	\label{fig:lattice}
\end{figure}

In this section we explore the distribution of segment ranks and edge ranks in a simple lattice model. In particular, we introduce a class of lattice models that we call ``Mondrian'' lattices (figure \ref{fig:lattice}), intended to mimic the growth process and the characteristics of hierarchical network-like patterns. In order to build the lattice, we start with a single rectangular domain and we iterate the following operation: each rectangular cell of the lattice is divided in two smaller rectangles by introducing a new cut parallel to one of its sides (let's say horizontal or vertical) at a random position chosen from a normal distribution around its centre. The cuts can either be ``alternating'', where horizontal cuts at time $t$ are followed by vertical cuts at time $t+1$, and vice-versa, or ``random'', where the horizontal or vertical direction of cut is chosen randomly for each cell and each division.

\begin{figure}[!b]
\vspace{0cm} 
\resizebox{1\columnwidth}{!}{
	\includegraphics{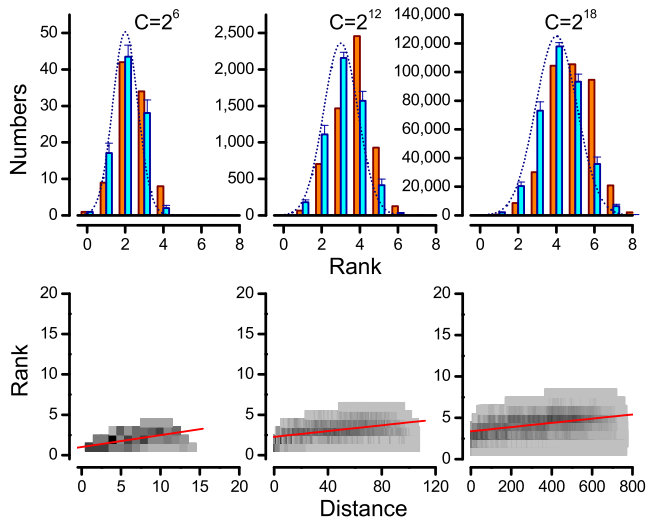}
}
	\caption{(Color online) Top row: histogram of segment ranks for Mondrian lattices of different size (from left to right the number of cells (C) is equal to $2^{6}$, $2^{12}$, and $2^{18}$). Orange (dark gray) histogram: alternating divisions. Cyan (light gray) histogram: random divisions (in the case of random divisions different realisations are possible: the whiskers on top of the histogram give the standard error of the mean). Bottom row: distribution of the rank vs. the topological distance of lattice edges for the same ``alternating'' Mondrian lattices. Red continuous line: linear fit to the data of the form $r(e_{i}) = \alpha \cdot td(e_{i}) + \beta$, with $td(e_{i})$ the topological distance of edge $e_{i}$ from the root edges. The slope ($tg(\alpha)$) is 0.148 for the lattice with $C=2^{6}$, 0.018 for the lattice with $C=2^{12}$  and 0.002 for the lattice with $C=2^{18}$.}
	\label{fig:lattices_6_12_18}
\end{figure}
Here, and in the rest of the paper, we will focus mainly on two kinds of statistics. The first is the histogram of \textit{segment} ranks (the frequency of occurrence of segments with a given rank). The second measure relies on \textit{edges} (not segments) and looks at how the ranks of individual edges increase with their topological distance from the roots. The topological distance between an edge $e_i$ and a root $e_r$ is the minimum number of junctions on a path between $e_r$ and $e_i$ in $G$. Intuitively, the rank measures the number of direction changes needed to reach a particular edge and the topological distance measures the number of vertices crossed. Hence, if we plot the rank $r(e_{i})$ of each network edge $e_{i}$ against its topological distance from the roots $td(e_{i})$, we obtain a characteristic distribution that can be fitted with a linear function of the form 

\begin{equation}
\label{eq:RvsDist}
r(e_{i}) = \alpha \cdot td(e_{i}) + \beta. 
\end{equation}

The slope of the fitting line is $tg(\alpha)$, and its inverse $\frac{1}{tg(\alpha)}$, gives the average length of the straight segments. This can also be interpreted as the largest scale at which spatial organization is observed.

Figure \ref{fig:lattices_6_12_18}, top row, displays the histogram of segment ranks for Mondrian lattices after 6, 12 and 18 iterations of divisions. The orange and cyan histograms are for alternating and random divisions, respectively. For random divisions the asymptotic distribution can be approximated by a Normal distribution:

\begin{equation}
N(r)=A\cdot e^{-\frac{(r-\overline{r})^2}{2\cdot\sigma^2}}
\label{theoreticalSegments}
\end{equation}

where $\overline{r} = 2 + \lfloor\frac{(t-2)}{6}\rfloor$ (where $t$ is the number of iterations of the division process), $A=\frac{2^{t+1}}{\sqrt{t+1}}\cdot 1.043$ and $\sigma = 1.020\cdot\frac{\sqrt{t+1}}{4}$. (Details on how the parameters $A$, $\overline{r}$ and $\sigma$ were estimated are provided in the appendix).
In practice, for a pattern observed at a given time, $t$ is usually unknown, and it is easier to approximate it in terms of the number of cells $C$, assuming that the relation $C=2^t$ holds . The number of cells for a planar graph in turn is obtained from the number of edges and vertices in the network through the Euler's formula, $V - E + C = 1$ (where $V$ is the number of vertices and $E$ the number of edges). In this way we can compute the theoretical rank distribution for lattices of given size. Such theoretical distributions are reported as dotted lines on top of the histograms of figure \ref{fig:lattices_6_12_18}, top row, as well as in different figures throughout the paper.

Figure \ref{fig:lattices_6_12_18}, bottom row, plots the rank $r(e_{i})$ vs. the topological distance $td(e_{i})$ for each network edge $e_{i}$ of alternating Mondrian lattices of different sizes (6, 12 and 18 iterations of division), together with a linear fit of the distribution (continuous red line). We can see that when the number of iterations of division increases, the overall slope decreases, revealing in fact a curved relation. This is not surprising, as topological distances typically scale with the square root of the lattice size, while ranks increase with the logarithm of the size.

\begin{figure}[t!]
\vspace{0cm} 
\resizebox{0.7\columnwidth}{!}{
	\includegraphics{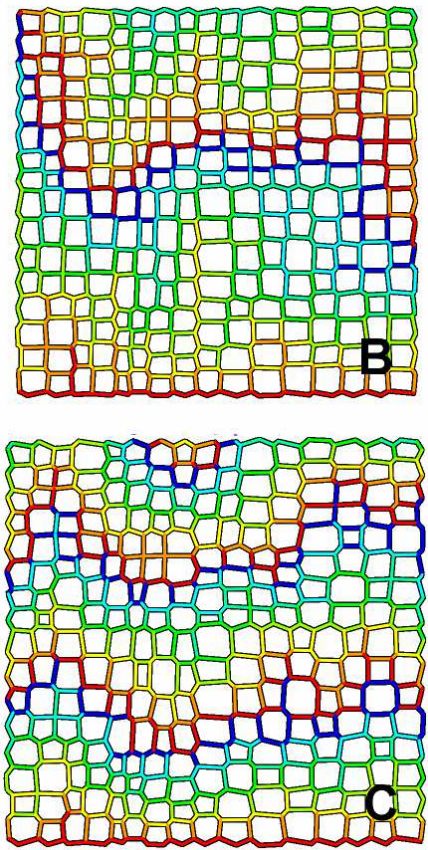}
}
	\caption{(Color online) The same lattice as in figure \ref{fig:lattice} where the position of each node is shifted by associating to all the adjacent edges a vector of unit length and centrifugal direction and computing the vectorial sum over all the adjacent edges. The pattern in B and C correspond to different numbers of iterations of degradation.}
	\label{fig:latticesdegradate}
\end{figure}
These simple lattice models may not be representative of the configurations of real world patterns with less regular edge orientations. In order to assess the robustness of rank indices when the edge orientations are altered, we proceeded as follows. Starting from a lattice of $2^{12}$ cells, obtained through alternating divisions, we iterated the following operation: at each time step the node positions are shifted by computing the resultant of vectors directed along the incident edges and oriented centrifugally. Each vector has magnitude arbitrarily fixed to $\frac{L}{30000}$, where $L$ is the length of the side of the grid. This progressively transforms the Mondrian lattice of figure \ref{fig:lattice}-A into the foam-like patterns of \ref{fig:latticesdegradate}-B and C. In order to prevent the lattice from shrinking during this process, we impose periodic boundary conditions by merging together the opposite sides of the lattice. After the degradation phase, the boundaries are separated again, and the rank computation proceeds normally, except that new nodes have been introduced along the perimeter of the lattice, doubling the nodes already existing on the opposite side of the lattice.


Such an evolution from orthogonal to symmetric junctions is not a purely geometric artifice: many real world patterns do actually evolve from an orthogonal to a symmetric configuration, as the result of forces that minimize the local length of the network edges. For instance, in basaltic lava rocks, the junctions, initially appearing at right angles, progressively evolve towards a symmetric configuration with angles of 120 degrees as the joints grow inward during solidification of lava~\cite{Aydin1988, Jagla2002}. In a similar way, leaf veins seem to intersect at right angles when they first appear, but then to evolve continuously towards more symmetric foam-like junctions~\cite{Bohn2002}. 

\begin{figure}[t!]
\vspace{0cm} 
\resizebox{1\columnwidth}{!}{
	\includegraphics{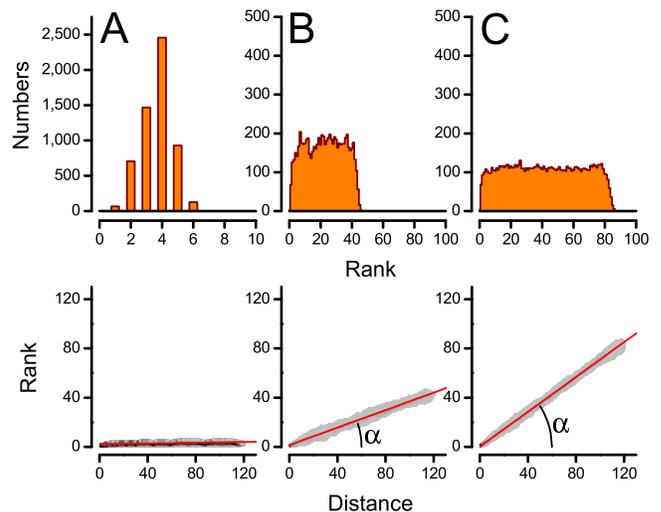}
}
	\caption{(Color online) From left to right: same plots as in figure \ref{fig:lattices_6_12_18} corresponding to three lattices with $2^{12}$ cells each, but different levels of degradation (0, 70 and 140 iterations of degradation respectively). When the local junction geometry is destroyed, the histogram of segment ranks (plots on the top) shifts from being peaked to uniform. The fitted slope of the rank vs. distance plot ($tg(\alpha)$) (bottom plots) is 0.014 for A 0.355 for B and 0.707 for C. Small differences between the graphs in A here and the corresponding ones in the middle column of figure \ref{fig:lattices_6_12_18} are due to the additional nodes introduced by imposing periodic boundary conditions during network degradation.}
	\label{fig:lattice12degradata}
\end{figure}
\begin{figure}[t!]
\vspace{0cm} 
\resizebox{0.6\columnwidth}{!}{
	\includegraphics{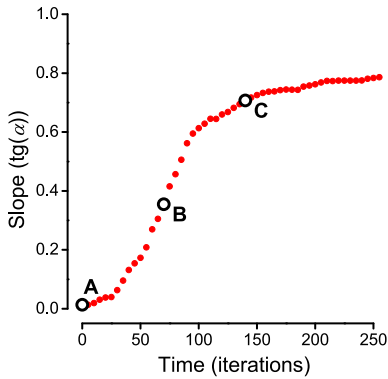}
}
	\caption{When the angles at junctions are altered by applying forces to the edges (as shown in figure \ref{fig:latticesdegradate}), the global organization of the pattern is rapidly lost. Correspondingly, the slope of the edge rank vs. distance ($tg(\alpha)$) increases towards 1. The plot reports this slope as a function of the number of iterations of degradations of the original orthogonal Mondrian lattice with $2^{12}$ cells. The three letters A, B and C mark the values of slope found for the three lattices of which figures \ref{fig:lattice} and \ref{fig:latticesdegradate} show a portion.}
	\label{fig:slopedegradation}
\end{figure}

Figure \ref{fig:latticesdegradate} shows two snapshots -at different steps of degradation- of the lattice in figure \ref{fig:lattice}. The three lattices (fig \ref{fig:lattice}-A and fig \ref{fig:latticesdegradate}-B,C) all correspond to a portion of the orthogonal Mondrian lattice with $2^{12}$ cells from which the plots of figure \ref{fig:lattice12degradata} are drawn. The plots in figure \ref{fig:lattice12degradata} report the statistics of segment histograms (top) and edge rank vs. edge distance (bottom) for the three levels of degradation of A (no degradation), B (70 iterations of degradation) and C (140 iterations). With increasing degradation, the histogram of segment ranks becomes flatter and its center shifts far to the right of the theoretical peak expected for pure Mondrian lattices.  The slope of edge rank vs. edge distance, that is nearly zero for the original lattice A, increases progressively. Figure \ref{fig:slopedegradation} shows that the slope undergoes a continuous transition on a relatively large scale (i.e. it is still relatively low also for networks, such as the lattice B, whose histogram of segment ranks is already flat). This means that once the large scale coherence is lost, the length over which coherence persists (inverse of the slope), decreases continuously. In other words the slope is a quantitative measure of the amount of disorder between the two extremes of a Mondrian lattice ($tg(\alpha) \approx 0$) and a foam like pattern ($tg(\alpha) \approx 1$). In the case of this latter, the rank increases at almost every junction.

In general, real world patterns are expected to deviate in various ways from these simple lattice models.  In the rest of the paper we explore the distribution of ranks and segment statistics in three different examples of real world network-like patterns: the pattern of fracture on the surface of materials, patterns of leaf veins in dicotyledon plants, and the pattern of urban streets in (unplanned) towns.

\section{Fracture patterns}
\label{sec:cracks}

Crack patterns often form on the surface of materials as the result of the shrinking of one material layer frustrated by its deposition on a non-shrinking substrate. This kind of pattern formation has been extensively observed and reproduced in controlled settings on a variety of materials, including mud, ceramics and coffee grounds.
The final patterns result from the combination of two distinct processes: the nucleation of new fractures and the propagation of already existing ones~\cite{Shorlin2000,Vogel2005,Toga2006}. Nucleation of new fractures usually involves the formation of tripartite junctions with equal angles of about 120 degrees~\cite{Toga2006}. Conversely, junctions formed by propagating fractures are the result of either two fractures meeting at a point (usually with an orthogonal angle), or of the branching of one growing fracture (in this latter case the angles formed at the junction are less predictable). If the crack pattern is produced in a non-elastic material, there is no reorganization after its formation and the final form of the pattern reflects the mechanisms of formation. Depending on the characteristics of the material, either nucleation of new fractures will be the most frequent process, or propagation of already formed fractures over long distances. Nucleation is more frequent in the case of very thin layers or inhomogeneous materials; propagation is predominant in brittle, homogeneous materials such as ceramics.

\begin{figure}[b!]
\vspace{0cm} 
\resizebox{0.8\columnwidth}{!}{
	\includegraphics{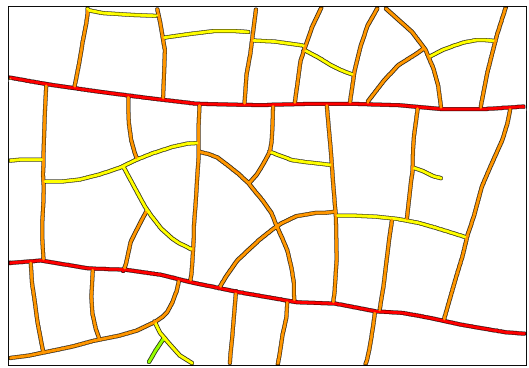}
}
	\caption{(Color online) Edge ranks computed for the fracture pattern of figure \ref{fig:examplecrack}. The root edges for rank computation are the two horizontal lines that cut the whole figure. The color coding is the same as for other figures. The $\square$ and $\bigcirc$ symbols are referred to in the main text.}
	\label{fig:excrackanalysed}
\end{figure}

\begin{figure*}[t!]
\subfigure{
\includegraphics[width=0.22\textwidth]{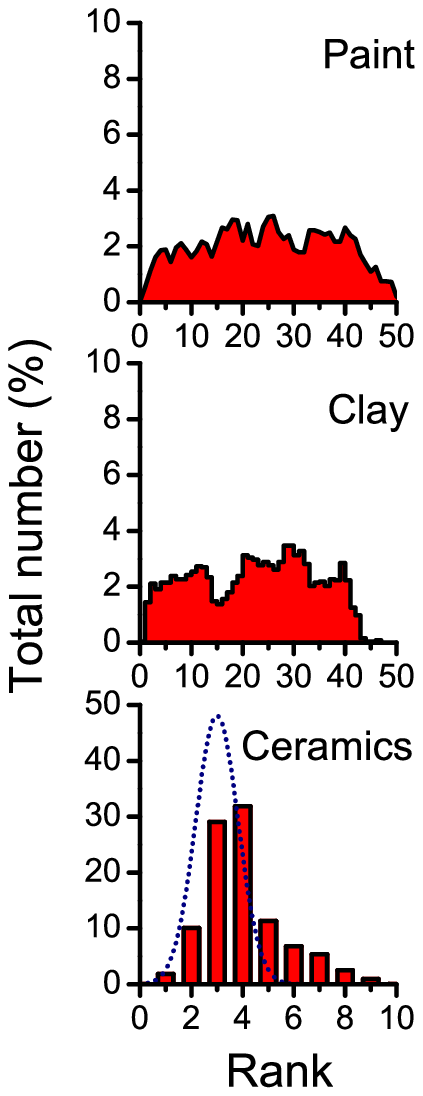}
}
\subfigure{
\includegraphics[width=0.22\textwidth]{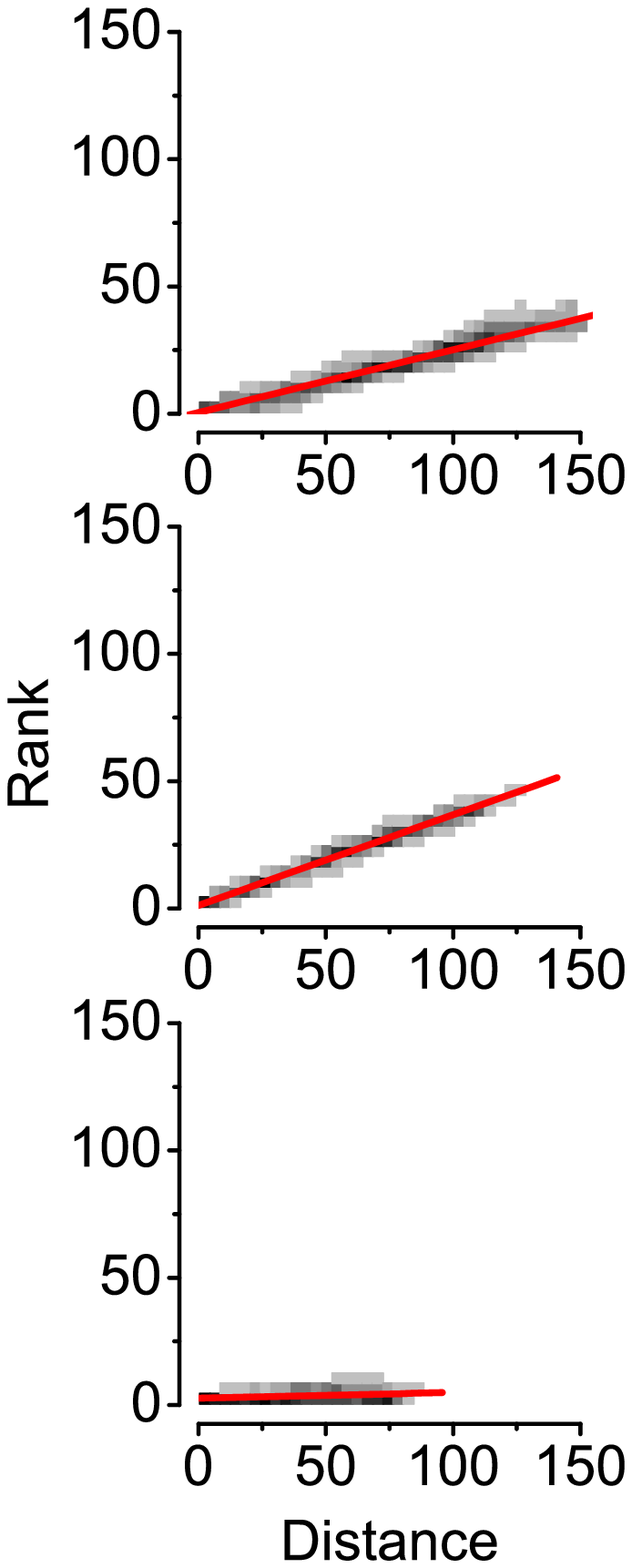}
}
\subfigure{
\includegraphics[width=0.45\textwidth]{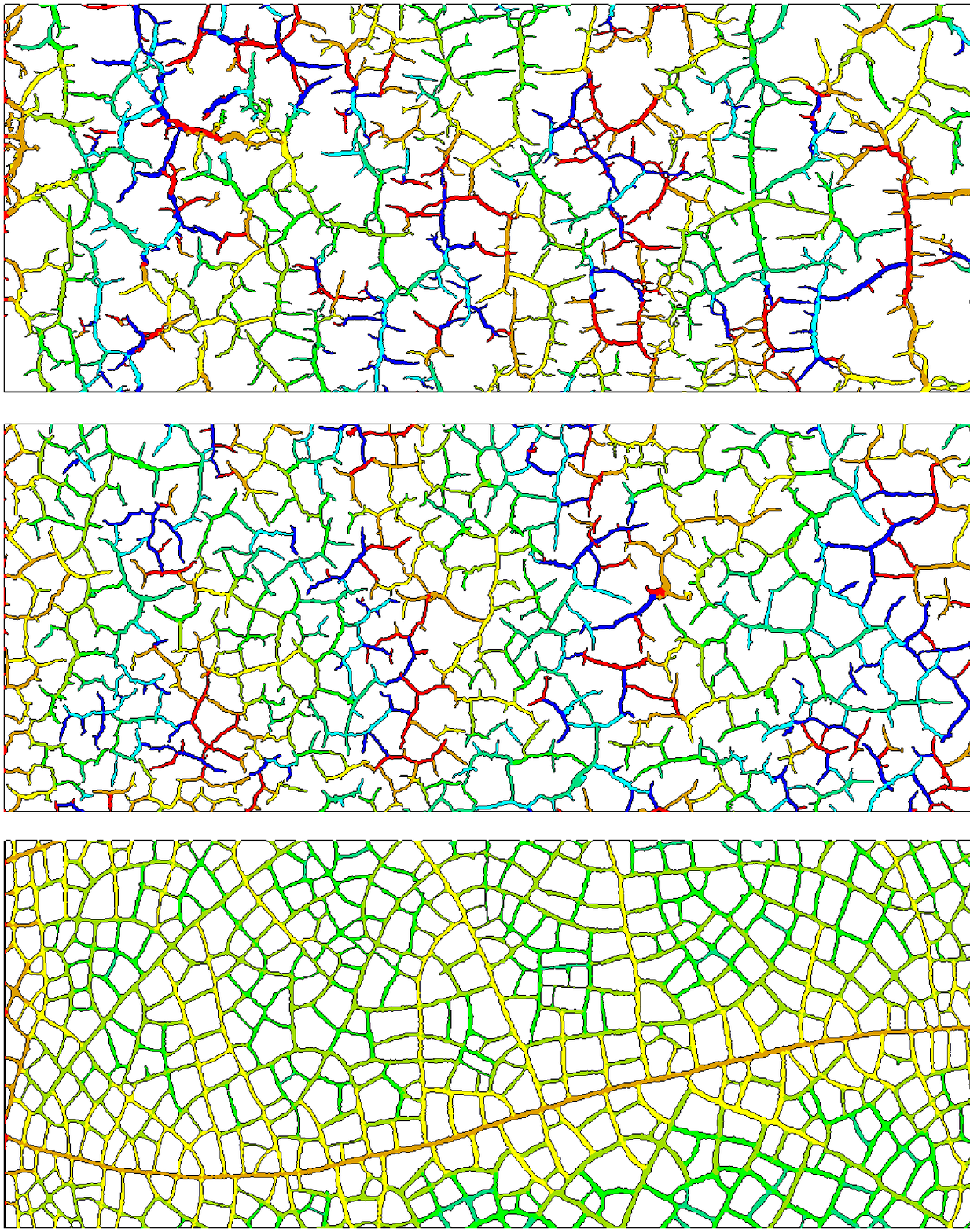}
}
\caption{Crack patterns formed in a layer of paint (top), in thin desiccating clay (middle), in the glaze of ceramics (bottom). The graphs on the left plot the histograms of segments with different ranks (as in previous figures); in the case of ceramics, we also plot the theoretical distribution for a Mondrian lattice of the same size. The graphs on the central column plot the ranks of edges vs. the topological distance of the same edges from the roots. The slope of best fitting straight line is 0.246 for paint, 0.356 for clay and 0.022 for ceramics. The images on the right are portions of the original patterns, where the fractures have been colored according to the rank of the corresponding network edge. The same periodic rainbow colormap as in figure \ref{fig:lattice} is used (color online). The roots for rank computation are all the edges on the left of the picture.}
\label{fig:cracks}
\end{figure*}
Patterns resulting mainly from the propagation of already formed fracture lines present a well defined hierarchy due to the sequential formation process. One such pattern is shown in figure \ref{fig:examplecrack}.  Figure \ref{fig:excrackanalysed} shows the ranks inferred by our algorithm when the two long horizontal lines are selected as roots. There is a partial mismatch between real and inferred hierarchy. This is in part due to the fact that when a cell is cut in two halves by a fracture, then the two halves become independent and it is no longer possible to establish a temporal relation between the new fracture events within a cell and those in the neighboring one. In the case of perfect hierarchical organization (as is the case for the pattern shown in figure \ref{fig:excrackanalysed}) one could also gain information by considering the organization of angles at both extremities of a segment: the fracture line marked with $\bigcirc$ terminates the fracture marked with $\square$, indicating that this latter is actually more recent. Unfortunately, a method that considers both extremities would lose generality and could not resolve a ``circular configuration'' where a segment ``A'' is terminated by another segment ``B'', the segment ``B'' is terminated by ``C'' and segment ``C'' is terminated by ``A''.

Once acknowledged that, depending on the characteristics of the material, some fracture patterns are mostly dominated by the nucleation of new fractures, and others by the elongation of existing ones, we want to test if our algorithm gives different classification results in the two cases. To this end, let us consider the crack patterns formed in three different materials: paint, desiccating clay and ceramics (figure \ref{fig:cracks} from top to bottom). The patterns were photographed with a digital camera, converted to grayscale images, high-pass filtered to remove inhomogeneities in the illumination, binarized by simple thresholding and cleaned applying a binary morphological majority filter (see~\cite{Gonzalez2002,Parker1997} for a review of common image processing techniques). The images of the fracture patterns were then skeletonized with a topology preserving algorithm based on distance transformation~\cite{Lee1994} to obtain a 8-connected skeleton (a skeleton where two pixels are considered to be connected if they share either a face or a corner, opposed to a 4-connected skeleton where only pixels that share a face are considered to be connected). For each skeleton pixel, we counted the number of pixels in their 8-neighborhood that also belonged to the skeleton, and all the pixels having a number of neighbors not equal to 2 were marked. All the connected sets of marked skeleton pixels were mapped into a network node. Whenever there was in the image an unmarked 8-connected path between two clusters of pixels identified as nodes we introduced an edge between the corresponding nodes. The orientation of each edge was estimated from the coordinates of the two endpoints. In a subsequent step, edges shorter than a threshold length (~3 pixels) were removed and the nodes at the two endpoints merged together.

\begin{figure*}[t!]
\vspace*{0cm} 
\resizebox{1\textwidth}{!}{
	\includegraphics{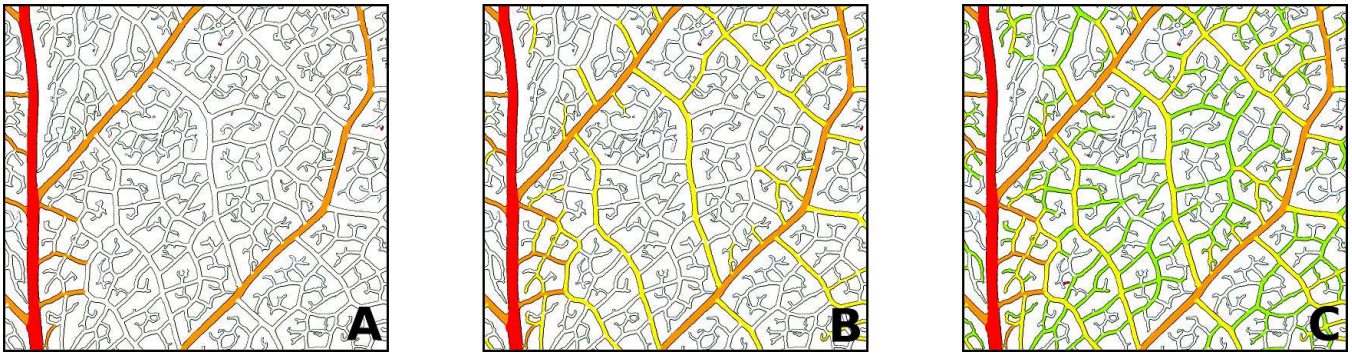}
	}
	\caption{(Color online) \textbf{A} Vein pattern of a dicotyledon: \emph{Hymenanthera chathamica}. \textbf{A}: only the veins with rank 0 and 1 are shown in colour (non white lines). \textbf{B}: only veins of rank 0,1 and 2 are colored. \textbf{C}: only veins of rank 0 to 3 are colored. (The whole leaf was analyzed, but for clarity only a small portion of the vein network is shown here).}
	\label{fig:leaves}
\end{figure*}
A rectangular section of the pattern is studied and all the edges crossing one side of the rectangle are selected as roots for the assignment of ranks. Figure \ref{fig:cracks} reports the statistics obtained on the three patterns, together with a snapshot of a small region of the original patterns (on the right), where the structure is colored according to the rank of the corresponding network edge.

Very different distributions are observed for the cracks formed in paint and clay, versus cracks formed in ceramics: in the former two, the lack of large scale organization is reflected into the linear increase of ranks with the topological distance from the root and in a flat histogram of segment ranks. Indeed, the junctions originating from nucleation of new fractures, whose angles are $\sim120$ degrees, always determine an increase in the rank of edges across the junction. The inverse of the slope of the curve fitting the data gives an indication of the length over which crack elongation proceeds: about three junctions. Conversely, for cracks formed in ceramics, the edge ranks are almost completely independent of the topological distance of edges from the roots, because of the large scale organization typical of these patterns. The slope of the rank vs. distance distribution (bottom plot, central column) is thus close to zero, but we can still see a deviation of the rank histogram toward the left of the theoretical normal distribution expected for a pure Mondrian lattice (Figure \ref{fig:cracks}, bottom left plot). Such small deviation is probably due to the fact that the analyzed region is a portion of a larger pattern and the cuts introduced by the arbitrary frame disconnected some edges from the network path that gives them their real rank. This also shows that rank distribution is a quite sensitive measurement. 

\section{Leaf venation networks}
\label{sec:leaves}

Leaf veins in the leaves of flowering plants form characteristic patterns that can be used by botanists as keys for taxonomic identification. However this identification is done by eye and does not rely on quantitative measurements. The pattern is hierarchical, and the diameter of a vein roughly reflects its order of appearance during leaf morphogenesis, with larger veins being older than smaller ones~\cite{Nelson1997}. Botanists define discrete vein orders looking at vein width at the point of branching from its parent vein: the large primary vein or midvein is continuous with the stem vascular bundles; secondary veins branch from the primary vein; tertiary veins are defined by their narrower width where they branch from the secondary veins and so on. 

We tested our ranking algorithm on the vein patterns of angiosperm leaves. The leaves were skeletonized with 10\% sodium hydroxide solution and the pattern was scanned with a commercial scanner in transmission mode with a resolution superior to 2000 pixels/inch and 256 gray levels. A network representation was extracted from the images in a similar way to what we described for cracks.
In the ranking procedure, the leaf stem is selected as root edge. This is in agreement with its special role for both transportation and leaf morphogenesis: the leaf stem is both the source (through xylem) and the sink (through phloem) of all transportation taking place in the leaf vein network, as well as the first vein to form during leaf morphogenesis~\cite{Taiz2010}.

Figure \ref{fig:leaves} shows a portion of leaf of \textit{Hymenanthera chathamica} with highlighted vein ranks obtained from our algorithm: veins with rank equal to 1 or 2 are shown in color in A, veins with rank 1 to 3 are highlighted in B and veins with rank 1 to 4 in C. The classification does not take into account the diameters of the veins, but only the orientation of the edges. Nevertheless, the results of classification match well the diameter of the veins: from the figure, we can see that veins marked with higher ranks have in general smaller size (which roughly corresponds to say that they have appeared later). In this sense, we also recover the vein patterns as derived from the botanical classification. This is consistent with the hypothesis that the older veins are not only larger but also straighter, as can be derived from the relation existing between vein sizes and junction angles~\cite{Bohn2002}. 
However, some small veins attached to the main vein are given a small rank, while one would intuitively ascribe them to a higher rank. This illustrates well the non complete reversibility of the fragmentation process: for each vein we can infer the order of appearance with respect to the parent vein, but not the exact age. Technically, also the botanical classification -based on diameters- has the same problem, because of the existing correlation between branching angles and diameter of veins.

\begin{figure}[t!]
\resizebox{1\columnwidth}{!}{
	\includegraphics{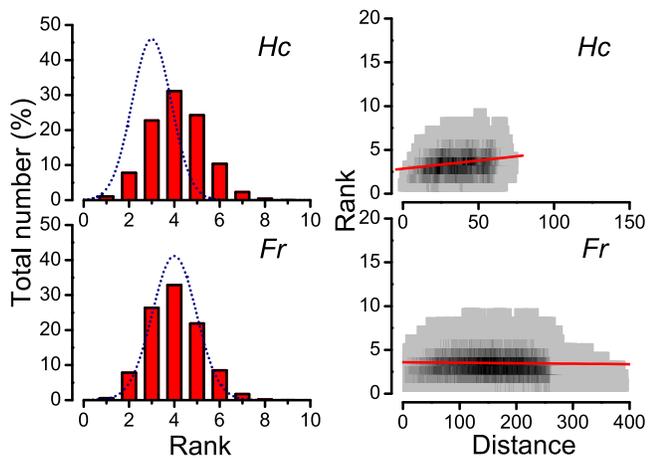}
}
	\caption{Top: Segement and edge statistics for \emph{Hymenanthera chathamica} ( \emph{Hc}, top) and \emph{Ficus religiosa} (\emph{Fr}, bottom). Network size is $\sim 12000$ nodes for \emph{Hc} and $\sim 130000$ for \emph{Fr}. The slope of the fit for the graphs on the right is 0.019 for \emph{Hc} and 0.000 For \emph{Fr}.}
	\label{fig:statsLeaves}
\end{figure}
Figure \ref{fig:statsLeaves} reports the histogram of segment ranks and the plot of edge rank vs. distance for both the \textit{Hymenanthera chathamica} (\textit{Hc}) leaf and for a leaf of \textit{Ficus religiosa} (\textit{Fr}). The \textit{Hc} network is about one magnitude order smaller than the \textit{Fr}. However, both networks are hierarchical: the histogram of segment ranks is peaked close to the theoretical value and the fitted slope of edge rank vs. distance is close to zero.

The larger \textit{Fr} pattern is very close to the perfect hierarchical pattern (its segment histogram overlaps quite well with the one of a perfect Mondrian lattice of the same size). Conversely, the histogram of segment ranks for \textit{Hc} deviates slightly from the theoretical one: it is still a Gaussian but shifted to higher values. Unlike the case of fractures, this cannot be ascribed to boundary effects as the whole leaf with the boundary veins is analysed, nor to a simple small number of veins, as the network is large and the distribution is already very well defined. This shift might be due to the different plasticity of the veins during their maturation in the two plant species. More measurements should be made to check if this shift is due to the difference in species or to the different maturation of the leaves depending on their sizes.   

\section{Urban street patterns}
\label{sec:streets}

\begin{figure}[b!]
\resizebox{0.8\columnwidth}{!}{
	\includegraphics{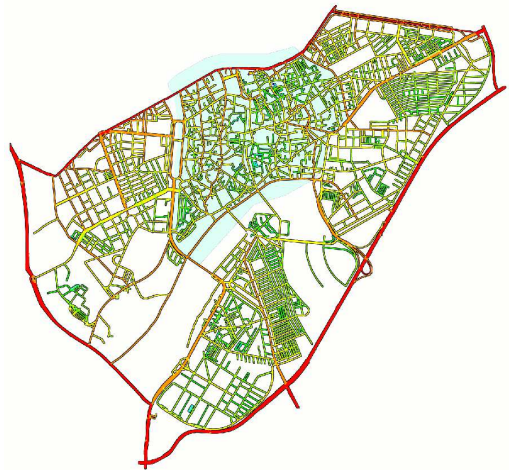}
}
	\caption{(Color online) Cordoba map, edges are colored according to their rank. }
	\label{fig:Cordoba}
\end{figure}
The growth of urban street systems is a complex process determined by a series of historical events and involving feed-back and regulation from global network processes, such as traffic and transportation. However, in a first approximation, for self-organized cities, the general form of these patterns can be described in terms of simple models where new streets appear over time with no reorganisation~\cite{Barthelemy2008,Barthelemy2009a,Courtat2010}. Within such models, the first streets connect the first houses to the country. As the urban pattern grows, new streets bifurcate from existing ones in the direction of not yet urbanized areas, or to join already existing streets. In this respect, the process of growth of urban streets shares similarities with the growth of fracture patterns~\cite{Bohn2005}, which supports the fact of assigning ranks to streets in a similar way.

We here test our ranking method on the towns of Cordoba (Spain) and Venice (Italy) and compare the statistics obtained for the two towns. In both cases we choose the perimeter of the town as root for the rank computation. (The shores of Venice and the highway ring around Cordoba).

\begin{figure}[t!]
\resizebox{1\columnwidth}{!}{
	\includegraphics{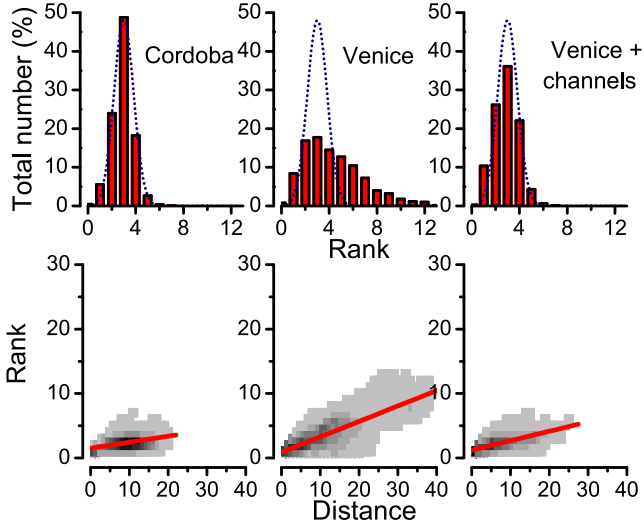}
}
	\caption{Top: Histogram of the number of segments in Cordoba street pattern, Venice street pattern and Venice street and channel pattern. Bottom: rank of network edges vs. their topological distance from the root edges. The slope of the fitted line to the bottom plots is 0.094 for Cordoba, 0.239 for Venice, and 0.146 when also the channels are considered in the network.}
	\label{fig:Towns}
\end{figure}

Figure \ref{fig:Cordoba} displays a map of the town of Cordoba, where each edge is colored according to its rank. The highlighted region in the figure corresponds to the historical city centre. Almost all the edges with highest rank appear to fall inside this region. Higher ranks often are the mark of lack of global organization. A possible outlook for future analysis could be testing if higher ranks correspond to parts of town that developed in a period of more self-organized, organic growth (e.g. periods when the central power was weaker). Overall, the histogram of segment ranks (reported in figure \ref{fig:Towns} top, left) is still compatible with that of a hierarchical network, that is, it overlaps with the distribution expected for a Mondrian lattice of the same size. The distribution of edge ranks vs. edge distance from the root edges is nearly flat, with a slight slope that can likely be ascribed to the relatively small size of the network.

When the same analysis is carried out on the street pattern of Venice, a quite different behavior is found: segments rank up to much higher values, suggesting an absence of large scale organization in this street pattern (figure \ref{fig:Towns}, middle column). This does not seem to be a pure artifact of our method, but agrees with other elements of the organization of Venice, where streets are not the principal element of organization of the town. This is reflected, for instance, in the fact that still today houses and buildings are not numbered according to their position along a street, but following a subdivision by districts (``sestieri''). 

\begin{figure}[t!]
\resizebox{0.8\columnwidth}{!}{
	\includegraphics{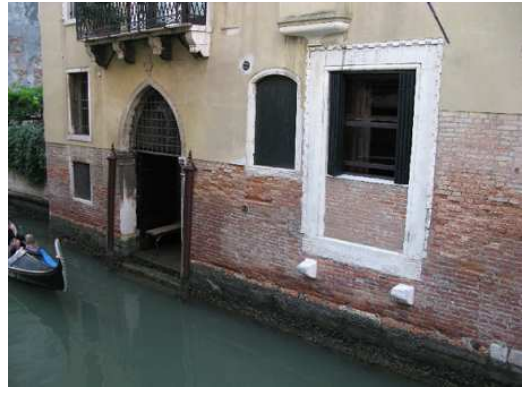}
}
	\caption{Direct access to the channel network from a house in Venice}
	\label{fig:Venezia}
\end{figure}

Interestingly, when for Venice one considers the transportation system including both streets and channels, the rank distribution gets closer to one of a hierarchical network (figure \ref{fig:Towns}, right). This reveals that the channels are the original part of the transportation system, and the origin of the town organization. The houses were first built along the channels with direct access to them (figure \ref{fig:Venezia}), and the streets appeared inside the islands delimited by the channels, as secondary divisions for inland house access. Thus removing the channels from the analysis is removing the large coherent structure,  and thus has a direct visible impact on the rank distribution. On the contrary, with the channels included, the hierarchical structure of successive divisions is recovered. 
So, the rank analysis has the potential of showing some peculiarities of specific towns, and at the same time of pointing to the possible reasons for these peculiarities.

\section{Conclusions}
\label{sec:conclusions}

In this article we introduced a new method to analyze spatial networks with loops and define quantitative measurements on them. The need to characterize quantitatively the form (and the possible morphogenesis) of network-like patterns motivated us to introduce a procedure for assigning ranks to all the edges of the pattern and group several edges into segments. Using a local spatial characteristic, the branching angles, we are able to quantify the large scale coherence of the pattern. 

The first measurement we propose, the distributionof segment ranks, is a very sensitive indicator of the existence of a large scale spatial coherence and of hierarchical subdivisions.  When this large scale coherence is lost, the measurement quantifies the scale over which organization persists, up to the point of purely local organization. This scale is directly given by the inverse of the slope of edge ranks vs. edge distances from the root. When tested on lattice models, these measurements show a dichotomy between two types of network, one with large scale coherence and one with purely local organization.

The method is efficient on different real patterns of various origin, from fractures to leaf venation and urban streets.  The obtained measures show the potential to discriminate between patterns with hierarchical history of growth and patterns grown out of more local rules. For fracture patterns, the two cases of many locally generated fractures that reconnect and few fractures propagating on large distance can be distinguished with no ambiguity. For leaf venation pattern, the method clearly reveals a hierarchical growth mechanism. In addition, the rank assigned to network segments correlates to some extent with the temporal order of appearance of the same segments, and hence the measure is informative on the process of growth itself. The matching between ranks and order of appearance however is not perfect, and the information provided by the ranks should be complemented from other sources for individual systems. For instance, with leaf veins one could consider both ranks and vein diameter to obtain a better classification of veins into discrete orders.

This analysis is also sensitive on town streets, and is able to reveal the particular organization of streets in Venice, whose structure can be explained as a secondary construction from the channels. The town structures are otherwise coherent with that of a sequential sub-division pattern, indicating some underlying logic in its development. 

We believe that the method can be generally applied to different types of patterns, revealing not only their structure but also in part the history of their growth. In the present paper, we built the classification from local angle information. More generally, however, any other local information can be used, such as -for instance- the size of the connecting element, to construct a similar hierarchy. It is thus a new, general and efficient way to  analyze networks with loops and group patterns of very diverse origins into the same structure and development classes.

\section{Appendix}
\label{sec:appendix}

\begin{figure}[b!]
\resizebox{0.9\columnwidth}{!}{
	\includegraphics{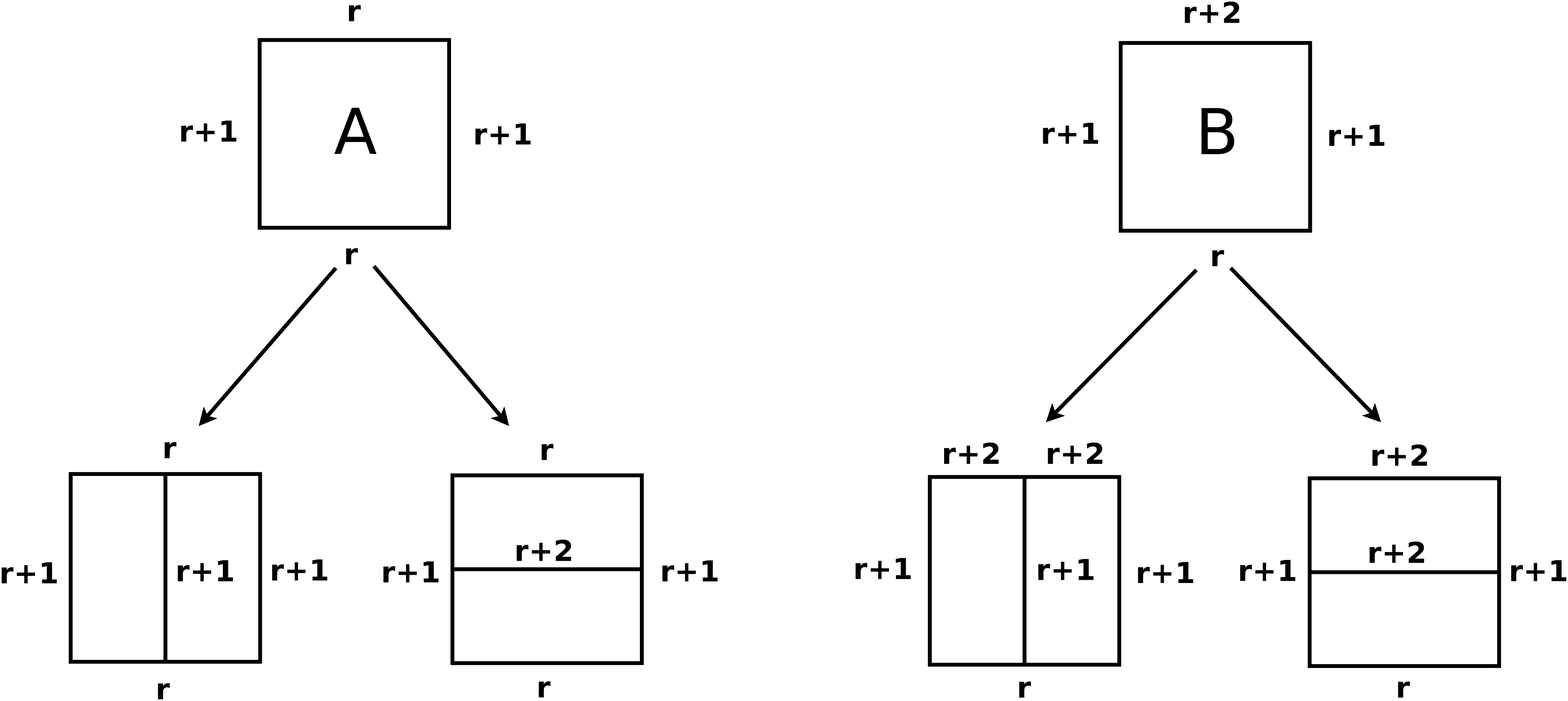}
}
	\caption{Schema of the possible cell types in a Mondrian lattice and of the results of their divisions. A cell is defined by the lowest rank ($r$) found on one of its sides and by the rank on the opposed side to this (the rank of the sides adjacent to that with rank $r$ is always $r+1$). The cell marked with the letter A has rank configuration $r, r+1, r, r+1$ over its perimeter and the one marked with B has configuration $r, r+1, r+2, r+1$. For each cell type the cut can be orthogonal to the side with rank $r$ or parallel to it.}
	\label{fig:diagram}
\end{figure}
The distribution of segment ranks in a Mondrian lattice can be computed more easily if one recognizes that the cuts within a cell only affect the distribution of segment ranks for that particular cell, but not for the neighbouring ones. We can identify two kinds of cells (``A'' and ``B'' in figure \ref{fig:diagram}), and two possible cuts for each cell (either orthogonal or parallel to the side with the lowest rank).
For any given cell and cut combination, both the resulting cells and the new segments added are determined: a cell of type ``A'', whose minimum rank is $r$ can give rise either to two cells of type ``A'' and with minimum rank $r$, or to two cells of type ``B'' also with minimum rank $r$. In the first case, the cut will introduce a new segment with rank $r+1$, while in the second it will introduce a segment with rank $r+2$. A cell of type ``B'' and minimum rank $r$ can generate two cells of type ``B'' and rank $r$; in this case it will add a new segment with rank $r+1$ as well as a new segment with rank $r+2$. This latter results from the ``cut'' operated by the new segment with rank $r+1$  on the already existing segment with rank $r+2$ (because we impose that a segment with higher rank is always ended when it meets one with lower rank). If the cut is in the other direction, the same ``B'' cell can generate one cell B with minimum rank $r$ and one cell ``A'' with minimum rank $r+1$.

If we assume that all the cuts have the same probability, that all the cells in the lattice are cut once for each time step and we take the average realizations, we obtain the following equations for the evolution of the pattern:
\begin{equation}
\begin{split}
A(r,t+1) = A(r,t) + \frac{1}{2}B(r-1,t)\\
B(r,t+1) = \frac{3}{2}B(r,t) + A(r,t)\\
Ns(r,t+1) = Ns(r,t) + B(r-2,t) + \frac{1}{2}A(r-2,t) +\\
+ \frac{1}{2}B(r-1,t) + \frac{1}{2}A(r-1,t)
\end{split}
\label{eq:segmentIterations}
\end{equation}
where $A(r,t)$ indicates the number of cells of type ``A'' with minimum rank $r$ at iteration $t$ and similarly for $B(r,t)$. $Ns(r,t)$ is the total number of segments with rank $r$ at iteration $t$.

\begin{figure}[t!]
\resizebox{0.9\columnwidth}{!}{
	\includegraphics{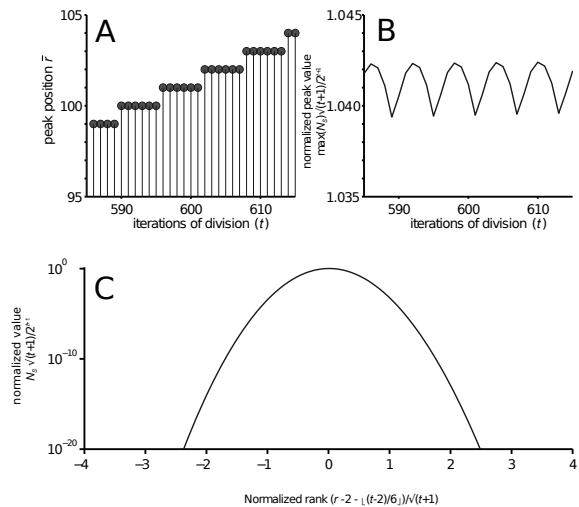}
}
	\caption{\textbf{A}. Plot of the position of the peak of the distribution of segment ranks $\overline{r}$ as a function of the number of iterations $t$ of equations \ref{eq:segmentIterations}. The plot is shown only for a few iterations of equations \ref{eq:segmentIterations} around $t=600$, but the staircase pattern is the same for all values of $t$. \textbf{B}. Plot of the rescaled maxima of the distribution of segment ranks as a function of the number of iterations of equations \ref{eq:segmentIterations}. \textbf{C}. The rescaled distribution after $t=1000$ iterations of equations \ref{eq:segmentIterations}.}
	\label{fig:Mondrian_iterations}
\end{figure}

We iterate the above equations, starting with a single cell of type $B$ and minimum rank $r=0$. With these formulae it is painless to go to very high numbers of cells and obtain well defined distributions. The first observation in the iterations is that the peaks of the distributions of $A$, $B$ and $Ns$ shifts to the right by one rank unit every six iteration steps. These shifts have different phases for $A$, $B$ and $Ns$. In particular the phase shift for $Ns$ is $t-2$, that is, the peak position $\overline{r}$ of $Ns$ increases by one unit with pace $\lfloor\frac{(t-2)}{6}\rfloor$. Figure \ref{fig:Mondrian_iterations}-A plots the position of the maximum $\overline{r}$ as a function of the number of iterations for a few iterations of equations \ref{eq:segmentIterations} around $t=600$.

We also guessed the way the maxima increase in amplitude with simple relationships: the scale of the numbers of cells $A$, cells $B$ and segments $Ns$ is by contruction $2^{t}$ (the number of cells). However, the maximum value does not simply scale in proportion to the number of cells, because the distribution gets larger while it moves to the right for increasing values of $t$. In particular, the scaling of standard deviation can be easily approximated by a simple formula as $\sigma=\sqrt{t+1}$. We found that if the the distribution is normalized by $\frac{2^{t+1}}{\sqrt{t+1}}$, then the maximum remains nearly stable through iterations (this is shown for a few iterations of equations \ref{eq:segmentIterations} around $t=600$ in figure \ref{fig:Mondrian_iterations}-B). 

When the distribution is rescaled in this way in width, position and amplitude, then it falls onto a perfect parabola in log coordinates (shown in figure \ref{fig:Mondrian_iterations}-C after 1000 iterations), without any change when increasing the number of iterations, except for attenuating fluctuations. So, even if we lack a theoretical demonstration, we know that the expression of $N$ in equation \ref{theoreticalSegments} is a Guassian and the dependencies on the parameters are asymptotically exact. Fitting a Gaussian on this distribution after a very large number of iterations ($t=1000$, that is, a number of cells of $2^{1000}$, -numerical approximation problems appear in our code around iteration 1024-), give the fitting parameters used in equation \ref{theoreticalSegments} of 1.043 and 1.020 respectively (these numbers could possibly be approximations of $\pi/3$ and $\sqrt{\pi/3}$, respectively).

\section{Acknowledgements}
\label{sec:acknowledgements}
We are grateful to Fabien Picarougne for comments and advice on the C++ implementation of the algorithm. We thank Alessio Cardillo, Christian Jost and Arianna Bottinelli for comments on previous versions of the manuscript.
This work was supported by ANR-06-BYOS-0008. A.P. was also supported by an ERC starting grant to David Sumpter (ref: IDCAB).

\end{document}